\documentclass{aastex631}
\usepackage{amsmath}

\begin{document}
\title{Navigating Unknowns: Deep learning robustness for gravitational wave signal reconstruction}

\tolerance=1
\emergencystretch=\maxdimen
\hyphenpenalty=10000
\hbadness=10000

\author[0000-0001-8700-3455]{Chayan Chatterjee}
\affiliation{Department of Physics and Astronomy, Vanderbilt University\\ 2201 West End Avenue, Nashville, Tennessee - 37235,}
\affiliation{Data Science Institute, Vanderbilt University\\ 1400 18th Avenue South Building, Suite 2000, Nashville, Tennessee - 37212, }

\author[0000-0003-1007-8912]{Karan Jani}
\affiliation{Department of Physics and Astronomy, Vanderbilt University\\ 2201 West End Avenue, Nashville, Tennessee - 37235,} 



\begin{abstract}

We present a rapid and reliable deep learning-based method for gravitational wave signal reconstruction from elusive, generic binary black hole mergers in LIGO data. We demonstrate that our model, \texttt{AWaRe}, effectively recovers gravitational waves with parameters it has not encountered during training. This includes features like higher black hole masses, additional harmonics, eccentricity, and varied waveform systematics, which introduce complex modulations in the waveform's amplitudes and phases. The accurate reconstructions of these unseen signal characteristics demonstrates \texttt{AWaRe}'s ability to handle complex features in the waveforms. By directly incorporating waveform reconstruction uncertainty estimation into the \texttt{AWaRe} framework, we show that for real gravitational wave events, the uncertainties in \texttt{AWaRe}'s reconstructions align closely with those achieved by benchmark algorithms like BayesWave and coherent WaveBurst. The robustness of our model to real gravitational wave events and its ability to extrapolate to unseen data open new avenues for investigations in various aspects of gravitational wave astrophysics and data analysis, including tests of General Relativity and the enhancement of current gravitational wave search methodologies.



\end{abstract}



\section{Introduction} \label{sec:intro}

Gravitational wave (GW) astronomy has undergone a remarkable evolution since the first detection in 2015, profoundly impacting our understanding of the universe's most violent and energetic phenomena \citep{GW150914}. The field now frequently reports new discoveries, signifying a major leap in our observational capabilities. The fourth observing run (O4) of the LIGO-Virgo-KAGRA Collaboration \citep{LIGO, Virgo, KAGRA} is currently in progress and plans are set for the next decade to further expand this network with both ground and space-based detectors \citep{LISA_1, LISA_2, LISA_IMBH_HM}. These advancements promise to increase the frequency of GW detections significantly, thereby enhancing our ability to conduct follow-up observations across various cosmic messengers, such as electromagnetic radiation and astrophysical neutrinos. This evolving landscape of GW astronomy sets the stage for significant advancements in understanding the universe through the lens of GW and their accompanying signals. \\

GW signals from merging compact binaries detected by the LIGO and Virgo interferometers are primarily identified using the template-based matched filtering technique \citep{Matched_filtering, FINDCHIRP}, which is most effective with stationary and Gaussian noise, but can be compromised by non-stationary noise transients in the detectors. Additionally, current low-latency matched filtering-based detection approaches often overlook the impact of precession \citep{Precession1, Precession2} in black hole spins and higher-order waveform harmonics, instead focusing only on the primary quadrupolar modes. This restriction limits our ability to explore significant astrophysical phenomena, such as the presence of intermediate-mass black hole (IMBH) binaries \citep{IMBH_HM} and hierarchical mergers, although recent work in these directions present significant advancements in the search for these elusive systems \citep{Portsmouth_precession, GstLAL_precession, Detection_HM_1, Detection_HM_2, Detection_HM_3}. Machine learning offers a transformative potential for GW data analysis, bringing sophisticated tools that can enhance detection capabilities and reduce computational burdens \citep{ML_MDC, ML_Review_paper}. In particular, deep learning models \citep{Deep_learning_Goodfellow} have been successfully applied to various aspects of GW astronomy, including rapid signal detection, noise characterization, and parameter estimation \citep{detection2, detection6, detection7, Schafer, Burst_ML_detection_1, MLy_pipeline, BNS_detection_ML, NSBH_detection_ML, DINGO_IS, Gabbard}.  \\

In \cite{AWaRe}, we demonstrated for the first time that deep learning can effectively reconstruct waveforms from precessing binary black hole (BBH) signals that include higher-order modes. This means given a segment of noisy detector strain data for such GW events, our model outputs the underlying astrophysical whitened waveform, removing it completely from background noise. Alternately, if the model is presented with a segment of pure noise data, with no underlying astrophysical signal, the model outputs a vector of tiny random fluctuations around zero amplitude, indicating that this method can be readily adapted into a rapid GW search algorithm. We will explore this application in a future work. Our model, \texttt{AWaRe}, or \textbf{A}ttention-boosted \textbf{Wa}veform \textbf{Re}construction network, evaluated using simulated signals embedded in actual LIGO Livingston noise from the third observation run (2019-2020) \citep{GWTC3}, consistently achieved a high overlap with the injected signals. This level of accuracy persisted across a broad spectrum of BH mass and spin configurations considered in our study. When applied to real GW events, our model's reconstructions achieved overlaps ranging from 85\% to 98\% with those obtained by Coherent WaveBurst (unmodeled) \citep{cWB, cWB_results} and LALInference (modeled) \citep{LALInference} methods, indicating that \texttt{AWaRe} is an effective and versatile tool for signal reconstrcution from a diverse range of compact binary systems. \\

One of the limitations of the approach presented in \cite{AWaRe} is that \texttt{AWaRe} was designed to only produce point-estimate predictions of the reconstructed BBH waveform, given a noisy strain time-series, without an estimate of the corresponding reconstruction uncertainty. This leads to biased estimates of the model's learning ability, since multiple waveforms can fit the same data, especially in the presence of noise and model limitations. Furthermore, having robust uncertainty estimates would enable a more rigorous validation of the model's generalization capabilities. Specifically, uncertainty quantification is critical for assessing whether \texttt{AWaRe} can accurately reconstruct waveforms it has not previously encountered, such as those from different waveform approximants, higher-order modes (HM), or scenarios with eccentric orbits. By providing confidence intervals or probability distributions around the predicted waveforms, we can determine the reliability of the model's predictions across these varied scenarios. This is essential for ensuring that the model not only performs well on familiar data but also adaptably handles the complex and diverse nature of real-world GW signals. \\

\begin{figure}[th!]
\centering
\includegraphics[scale=0.36,trim = {30 0 0 0}]{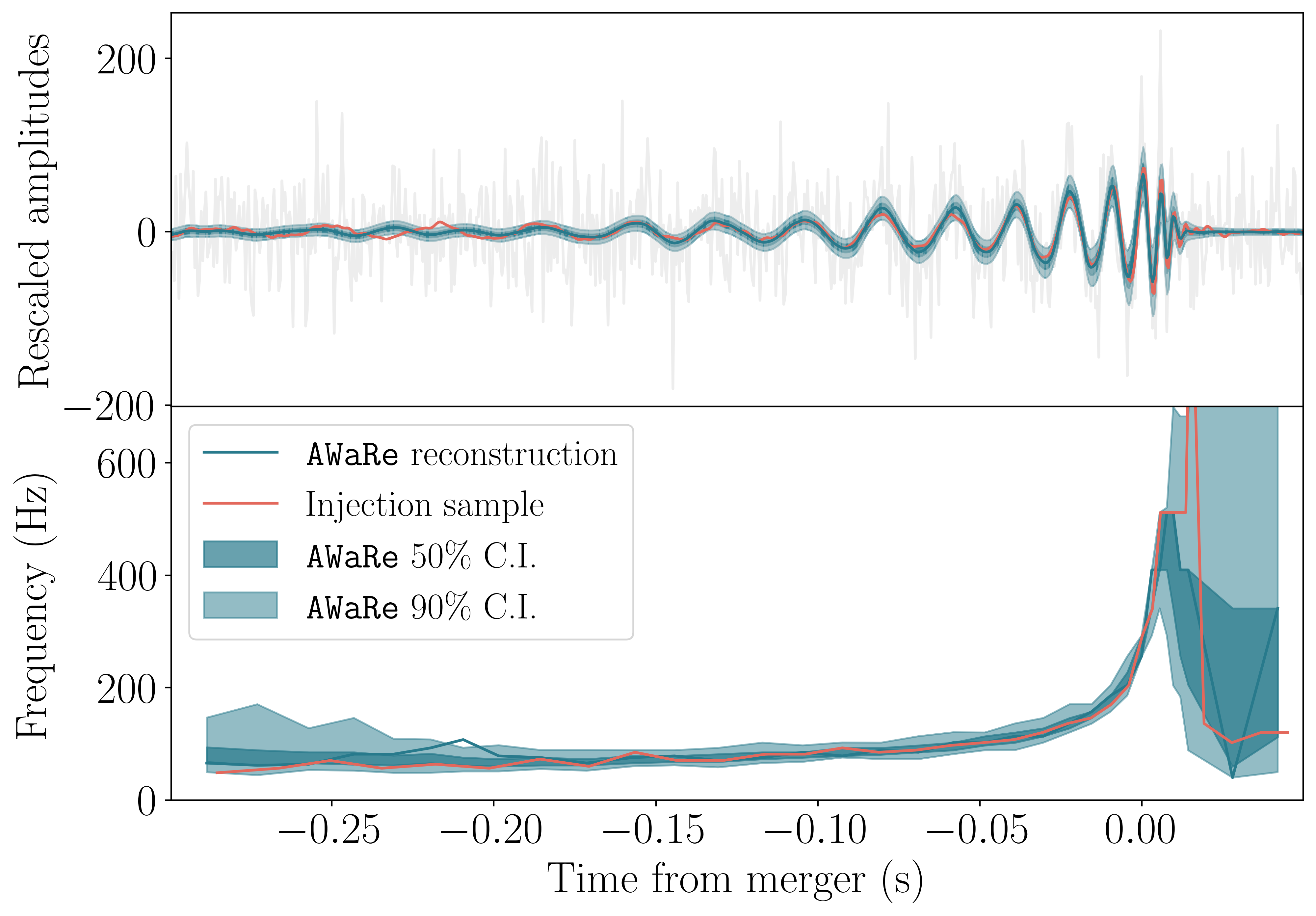}
\caption{Top: Reconstructed signal (blue) corresponding to the injected waveform (red) plotted against the whitened strain
(grey), obtained by injecting the simulated waveform in LIGO O3 noise. The dark and light blue regions show 50\% and 90\%
confidence intervals of the reconstructed waveform. Bottom: The frequency evolution over time of the waveform and its reconstruction shown in the figure on top.}
\label{fig:Example}
\end{figure}
In this article, we present an enhanced version of \texttt{AWaRe}, introduced in \cite{AWaRe}, in which we incorporate uncertainty estimation into the GW waveform reconstruction. We also perform tests of the generalizability of the model for different types of BBH sources. The remainder of this paper is organized as follows: Section 2 describes the modifications made to our model from \cite{AWaRe} for uncertainty estimation. Section 3.1 demonstrates \texttt{AWaRe}'s performance on real GW events, along with a comparison against BayesWave (BW) \citep{BayesWave, BayesWave1} and cWB algorithms. In Sections 3.2 -- 3.3, we discuss our model's robustness on injections in real LIGO Livingston noise for signals whose parameters lie outside our training set distributions, including different waveform families, higher primary masses and non-zero eccentricities. Finally, in Section 4, we discuss the implications of this work and the applicability of this model for GW detection and detector characterization. \\

\begin{figure*}[th!]
\includegraphics[scale=0.87,trim = {10 0 0 0}]{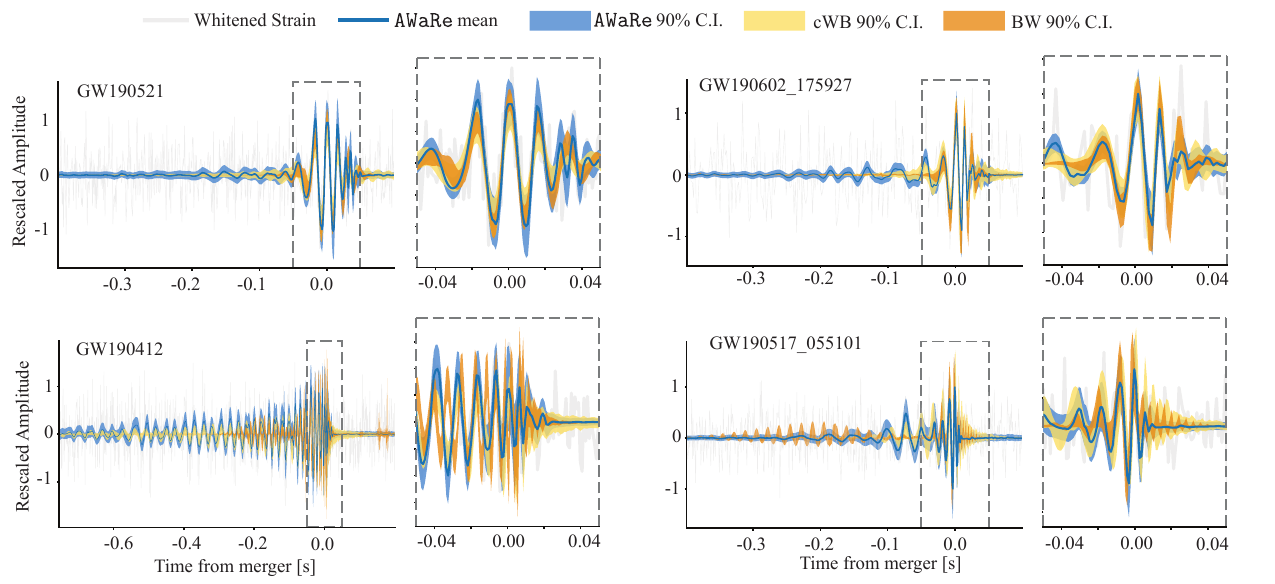}
\caption{\label{fig:Real_events} Clockwise from top left: \texttt{AWaRe} mean reconstruction and 90\% C.I. (blue), cWB 90\% C.I. (yellow) and BW 90\% C.I. (orange) for GW190521 ($\mathcal{M_{\text{det}}}$ = 113.48 M$_{\odot}$), GW190602\_175927 ($\mathcal{M_{\text{det}}}$ = 72.18 M$_{\odot}$), GW190412 ($\mathcal{M_{\text{det}}}$ = 15.30 M$_{\odot}$) and (d) GW190517\_055101 ($\mathcal{M_{\text{det}}}$ = 35.68 M$_{\odot}$). The numbers in the brackets refer to the median detector frame chirp mass of these events \citep{GWTC2}. The grey lines show the whitened strain data provided as input to the model.}
\end{figure*}

\section{Method}

Our model, \texttt{AWaRe} is a deep learning model consisting of two neural networks -- an encoder, consisting of one-dimensional convolutional layers \citep{CNN_1, CNN_2} that extract features from noisy GW strain data, and a decoder, consisting of Long Short-Term Memory (LSTM) \citep{LSTM} layers that reconstruct the pure waveforms from the embeddings learned by the encoder. In addition to the encoder and decoder layers, we apply a multi-headed attention (MHA) mechanism \citep{Attention} on top of the encoder stack which generates a richer feature representation from the input, allowing the model to better focus on relevant patterns in the data, thereby enhancing the accuracy and efficiency of the waveform reconstruction. This integration of MHA provides significant improvements in handling complex waveform features and also makes \texttt{AWaRe} particularly effective in distinguishing genuine signals from noise. Section 2.2 in \cite{AWaRe} provides a detailed description of the \texttt{AWaRe} model architecture and the MHA algorithm. \\

In the present work, we enhance the predictions of \texttt{AWaRe} by introducing the estimation of waveform reconstruction uncertainty. To achieve this, we assume a Gaussian distribution for every sample of our model's prediction and train our network to output the mean and standard deviation for these Gaussian distributions. In particular, for a 1-second long input strain data sampled at 2048 Hz, our model predicts 2048 means and standard deviations, corresponding to the 2048 independent Gaussian random variables used to parameterize the reconstructed waveform. These mean and standard deviation values are then used to compute the 90\% confidence intervals of the model's prediction of the recovered waveforms. Fig.~\ref{fig:Example} shows the amplitude and frequency evolution of a 30 + 30 M$_{\odot}$ GW signal (red) and its reconstruction with associated uncertainty (blue) obtained using \texttt{AWaRe}. To obtain the frequency evolution plot, the method described in Sec. V B in \citep{BayesWave1} was used.



\section{Results}

We present results obtained by \texttt{AWaRe} on real GW events detected during the third observation run (O3) \citep{GWTC3}, as well as on simulated injections with source parameters beyond the chosen prior distributions of our training set. To characterize the quality of waveform reconstruction, we compute the mismatch ($\mathcal{M}$) between \texttt{AWaRe}'s mean reconstruction and the maximum likelihood waveform reconstruction from cWB and BW, for real events, and the injected waveforms for simulated injections. \\

The mismatch quantifies the dissimilarity between two waveforms and is defined as,

\begin{equation}
\mathcal{M}(h, s) = 1 - \max_{t_{c},\phi_{c}}\mathcal{O}(h, s),
\end{equation}

where $\max_{t_{c},\phi_{c}}\mathcal{O}(h, s)$ represents the maximum overlap, as defined in Eq. 7 in \citep{AWaRe}, maximized over the time ($t_c$) and phase ($\phi_c$) of coalescence. Here, $h$ may refer to the reconstructed waveform produced by \texttt{AWaRe}, and $s$ refers either to the injection waveform or to the maximum likelihood reconstruction obtained by cWB or BW for real events. A higher value of $\mathcal{M} $ denotes higher dissimilarity between two waveforms. The mismatch values quoted throughout the paper refers to the mismatch computed between the mean reconstruction obtained from \texttt{AWaRe} (waveform obtained by taking the mean value of the 2048 independent Gaussians) and the injection sample.

\subsection{Tests on real O3 events}

We tested \texttt{AWaRe}'s performance on real O3 events and compared the reconstruction with results from cWB and BW. Figures~\ref{fig:Real_events} shows the mean \texttt{AWaRe} reconstruction (in blue) and 90\% confidence interval (in light blue) against the 90\% confidence intervals from cWB (in yellow) and BW (in orange). The grey curves across all the plots show the whitened strain data provided as input to the model. \\

The top left plot shows our result for GW190521, an interesting candidate due to its high total mass and support for misaligned spins (Bayes factor = 11.5) \citep{GW190521}. There is also strong observational evidence for a multimode BH ringdown spectrum \citep{GW190521_HM}. Our findings indicate that the mean \texttt{AWaRe} reconstruction lies well within the 90\% C.I. of both cWB and BW, and the 90\% C.I. also overlaps well with that of the cWB and BW 90\% C.I. However, \texttt{AWaRe} predicts an additional cycle at the end of the merger region, which is not present in the other two analyses. \\

GW190602\_175927 (top right, Fig.~\ref{fig:Real_events}) is another interesting GW candidate due to its high total mass. Our results again demonstrate very high reconstruction accuracy relative to cWB and BW around the merger, with small additional fluctuations in the inspiral phase that are not observed in the other analyses. The high sensitivity of \texttt{AWaRe} towards heavy binaries and burst-like real sources motivates applying this method for the analysis of IMBH binaries. We will investigate this in our future work. \\

\begin{figure}[t!]
\centering
\includegraphics[scale=0.35,trim = {30 0 0 0}]{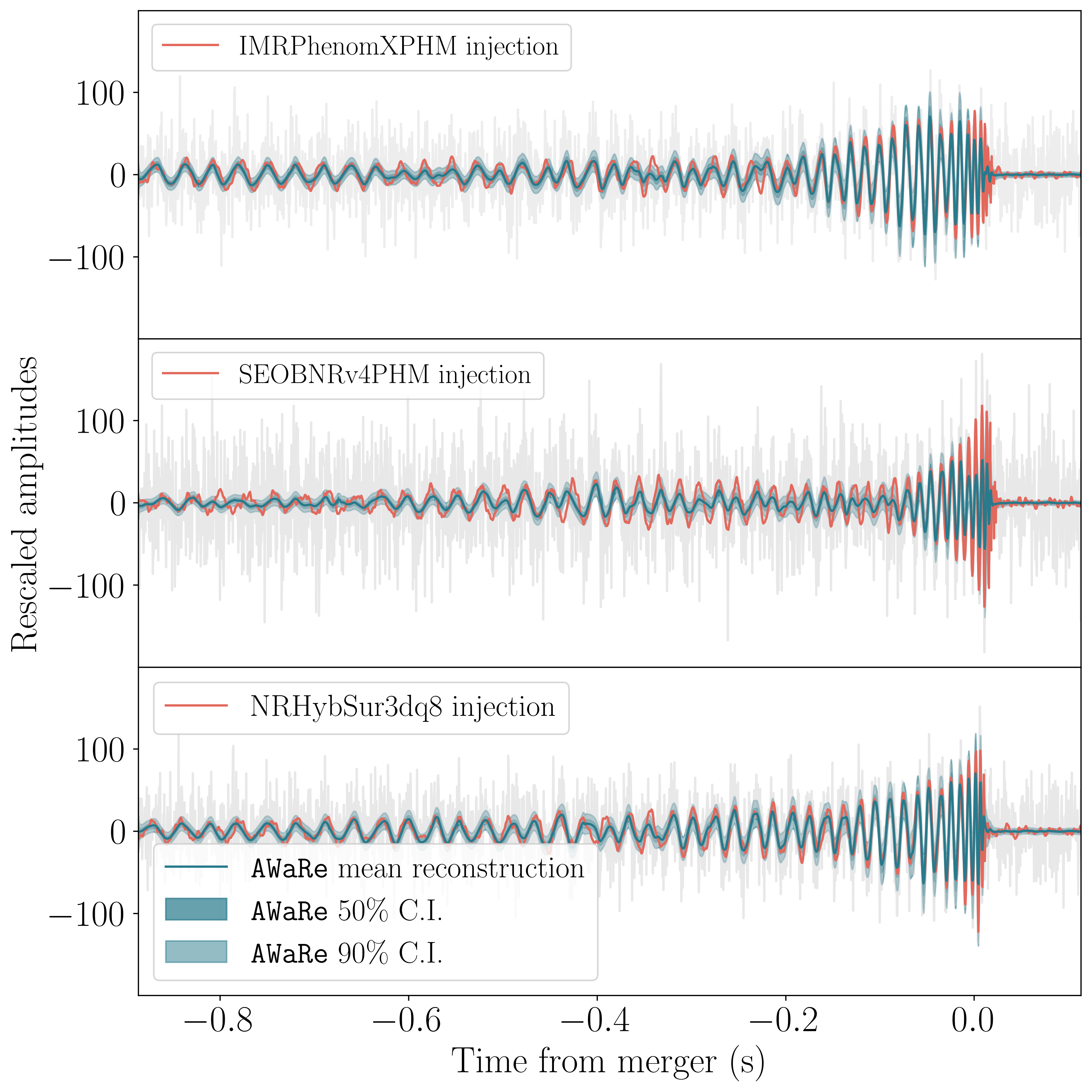}
\caption{\texttt{AWaRe} mean reconstruction with 50\% and 90\% C.I. (blue) for GW190412-like injection (red) generated using (a) IMRPhenomXPHM (b) SEOBNRv4PHM and (c) NRHybSur3dq8 waveform approximants. The grey lines show the whitened strain data.}
\label{fig:GW190412_waveforms}
\end{figure}

The bottom left figure in Fig.~\ref{fig:Real_events} shows our reconstruction for GW190412, a notable case due to its higher order mode contributions \citep{GW190412}. The event's unequal masses ($q = m_{2}/m_{1} = 0.28^{+0.12}_{-0.07}$), non-zero $\chi_\text{{eff}}$ ($0.25^{+0.08}_{-0.11}$), and $\chi_{p}$ ($0.31^{+0.19}_{-0.16}$) make it an ideal test for our model. \texttt{AWaRe}'s mean reconstruction and 90\% confidence interval (C.I.) align in phase with the cWB 90\% C.I., though \texttt{AWaRe} shows greater uncertainty during the inspiral. The BW 90\% C.I., however, exhibits significant mismatches in amplitude and phase compared to both \texttt{AWaRe} and cWB, predicting more oscillations. This discrepancy may arise because both cWB and BW use wavelets optimized for burst-like sources, not low chirp-mass events like GW190412. BW uses wavelets within a Bayesian framework for precise reconstruction \citep{BayesWave, BayesWave1, BayesWave2}, while cWB employs wavelets for time-frequency analysis and coherent detection across detectors, focusing on unmodeled burst signals \citep{cWB}. Since these methods do not rely on modelled template banks for searches and reconstruction, they are most effective for analysis of burst-like GW sources, corresponding to events with high total masses ($> $ 60 M$_{\odot}$). \\

\begin{figure*}[t!]
\includegraphics[scale=0.82]{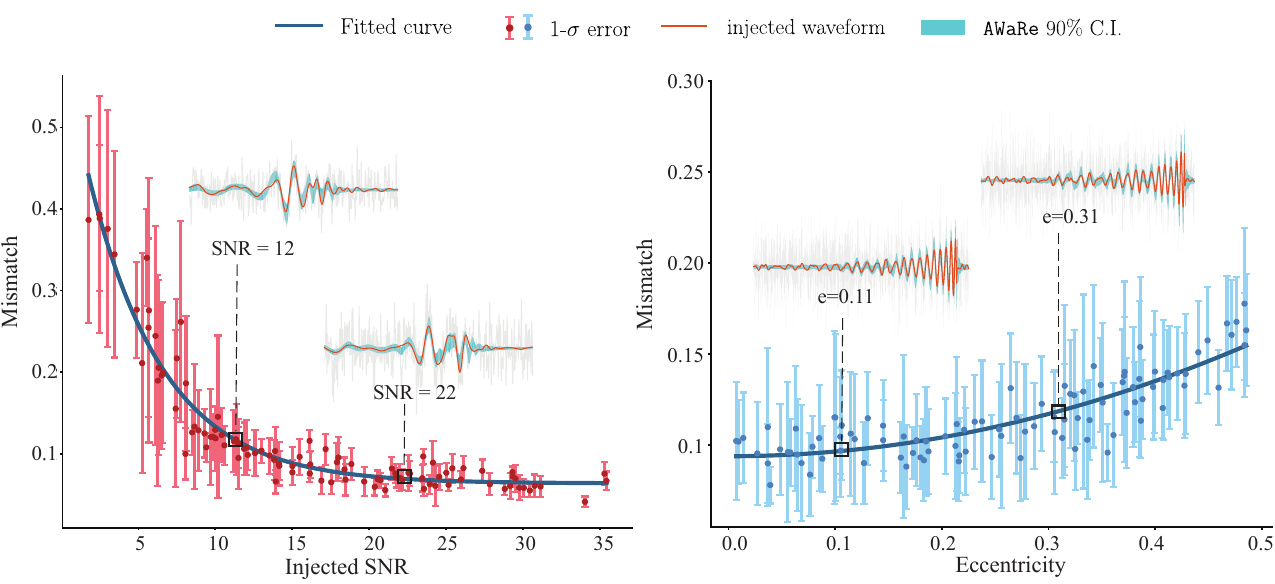}
\caption{(Left) Mismatch between \texttt{AWaRe} mean reconstructions and pure waveforms as a function of SNR for injections in LIGO O3b noise. (Right) Mismatch between \texttt{AWaRe} mean reconstructions and pure waveforms as a function of eccentricity for injections in LIGO O3b noise.
The red dots (left) and blue dots (right) and associated error bars show the mean mismatches and the 1-$\sigma$ uncertainties obtained by running \texttt{AWaRe} on injections in 10 different realizations of O3b noise. The dark blue lines shows the best fit curves. The inset plots shows pure waveforms (red) and 90\% C.I. of \texttt{AWaRe} reconstruction (light blue) for a SNR-12 and SNR-22 sample respectively (left) and waveforms with eccentricities of 0.11 and 0.37 respectively (right).}
\label{fig:Convergence_test}
\end{figure*}

The bottom right figure in Fig.~\ref{fig:Real_events} shows result for GW190517\_055101, which is a more generic BBH signal with primary and secondary masses, $37.4^{+11.7}_{-7.6}$ and $25.3^{+7.0}_{-7.3} $ respectively \citep{GWTC2} in the source frame. The result shows better match with cWB and BW than GW190412, because of its relatively higher masses for the reason described earlier. Our testing of \texttt{AWaRe} on real O3 events, compared with cWB and BW, demonstrates its robust performance. The framework's ability to handle a diverse range of events underscores its potential for broader application in GW detection and detector characterization.


\subsection{Tests on different waveform approximants}

In this study, we demonstrate the performance of \texttt{AWaRe}, trained exclusively on IMRPhenomXPHM waveforms, in reconstructing GW signals from three different waveform families: IMRPhenomXPHM \citep{IMRPhenomXPHM}, SEOBNRv4PHM \citep{SEOBNRv4PHM}, and NRHybSur3dq8 \citep{NRHybSur3dq8}. Since these waveform families significantly differ in their higher order mode content, this test serves as a strong indicator of the generalizability of our model to complex waveform systematics outside our training set. \\

The top, middle, and bottom panels in Fig.~\ref{fig:GW190412_waveforms} show the whitened strain data, the respective injected waveforms from each family, and the mean reconstructions by \texttt{AWaRe} for a GW190412-like injection sample \citep{GW190412}. The \texttt{AWaRe} reconstructions are shown in blue and the injected waveforms are shown in red across all three waveform families. GW190412 was chosen in particular because of the statistical evidence for the presence of higher order modes, as previously discussed \citep{GW190412}. We performed the injections with SNR of 22, which is higher than the observed network SNR of $19.0^{+0.2}_{-0.3}$ \citep{GW190412} in order to enhance the effect of the available higher harmonics in the three waveform families. The shaded regions represent the 50\% and 90\% confidence intervals (C.I.) of the reconstructions. \\

For IMRPhenomXPHM, the model displays high reconstruction accuracy ($\mathcal{O}$ = 0.88 with injected waveform), which is expected, by virtue of the fact that \texttt{AWaRe} is trained on IMRPhenomXPHM injections. For SEOBNRv4PHM ($\mathcal{O}$ = 0.71 with injected waveform), the amplitude near the merger is underestimated, which could be attributed to the discrepancy in the higher order mode content, which has its strongest influence where the signal is loudest, i.e., near merger. The reconstruction of NRHybSur3dq8 ($\mathcal{O}$ = 0.87 with injected waveform) shows the full injection waveform is captured within the 90\% \texttt{AWaRe} bounds, and the mean reconstruction also closely follow the injected waveforms indicating high reconstruction accuracy. These results imply that the model's training on a single waveform family (IMRPhenomXPHM) generalizes well to other families, enhancing its robustness and generalizability, which is crucial for improving GW detection in various astrophysical scenarios.

\subsection{Tests on masses beyond the training set}

In order to further test the generalizability of our model, we tested \texttt{AWaRe}'s performance on injections with primary mass beyond the training set. \texttt{AWaRe} was initially trained using IMRPhenomXPHM waveform injections with component masses $m_{1}$ and $m_{2}$ uniformly distributed in the range [5, 100] M$_{\odot}$ in the detector frame. To evaluate the model's robustness, we tested it on injections with primary mass, m$_{1}$ 
uniformly distributed in the range 200 and 250 M$_{\odot}$. In addition, we chose the mass ratio to be strictly $>$ 4, and the inclination angle was uniformly sampled between 75 and 105 degrees. These parameters were chosen to enhance the effect of higher-order modes in the signals, providing a rigorous test of the model's capabilities to reconstruct complex waveform systematics in a parameter space outside the scope of our model's training. \\

The left plot in Fig.~\ref{fig:Convergence_test} presents the mismatch between the reconstructed waveforms and the injected waveforms as a function of the injected SNR. The primary y-axis on the left shows the mismatch values, while the x-axis represents the injected SNR. The red dots correspond to the mean mismatch values, with error bars representing the 1-$\sigma$ uncertainty. The blue line indicates a fitted curve to the data points. To obtain the points and the error bars in the mismatch plot, we injected the same waveform in 10 different realizations of real LIGO O3b noise and then computed the mean and standard deviation of the mismatches between the mean reconstructions from \texttt{AWaRe} and the injected waveform. The main plot shows that the mismatch decreases as the injected SNR increases, demonstrating that \texttt{AWaRe} performs better with higher SNR signals. At lower SNRs, the mismatch values are higher and exhibit larger error bars, indicating greater variability and less accurate reconstructions. As the SNR increases beyond approximately 15, the mismatch values stabilize and the error bars become smaller, suggesting more consistent and reliable reconstructions. \\

The insets provide a detailed view of the time-domain reconstructions. The top inset corresponds to an injected waveform with SNR - 12, while the bottom inset shows a waveform with SNR - 22. In both cases, the gray lines represent the whitened strain data, the red lines represent the whitened injected waveforms, and the light blue bands show the 90\% C.I. of the reconstructions. These insets illustrate that the \texttt{AWaRe} can closely follow the injected waveforms with the confidence intervals adequately capturing the uncertainty in the reconstructions. \\



\subsection{Tests on eccentric waveforms}

In this study, we extend the evaluation of \texttt{AWaRe} to reconstruct eccentric waveform injections generated with the EccentricTD waveform approximant \citep{EccentricTD}. The injected signals have component masses of 30 and 30 M$_{\odot}$, are spinless, and have a SNR of 12. Notably, all waveforms in the training set had zero eccentricity, providing a stringent test of the model's ability to generalize to waveforms with non-zero eccentricity, which introduce modulations in amplitude and phase. The reconstruction of signals with non-zero eccentricity using \texttt{AWaRe} has important implications in the future application of our method for detecting eccentric BBH merger signals. This work would add to recent works in the reconstruction and detection of eccentric GW signals using modelled, unmodelled and semi-modelled methods \citep{eccentricity_search, Soumen_Roy_reconstruction}.\\

The right plot of Fig.~\ref{fig:Convergence_test} illustrates the mismatch between the reconstructed waveforms and the injected waveforms as a function of the initial eccentricity (i.e., eccentricity at 20 Hz). The y-axis shows the mismatch values, while the x-axis represents the initial eccentricity. Blue dots indicate the mean mismatch values, with error bars representing the 1-$\sigma$ errors. The dark blue line shows a fitted curve to the data points. To obtain the data points and error bars, we injected the same waveform in 10 different realizations of real LIGO O3b noise and then computed the mean and standard deviation as described in the previous section. The plot reveals that while the mismatch increases with higher eccentricities, indicating the greater challenge in accurately reconstructing waveforms with significant eccentricity, \texttt{AWaRe} nonetheless shows remarkable generalization capabilities. At lower eccentricities, the mismatch values are lower with smaller error bars, indicating more accurate and consistent reconstructions. Even at higher eccentricities, although the mismatch values increase, the model still manages to produce reliable reconstructions. \\

The insets provide a closer look at the time-domain reconstructions for two specific eccentricities. The top inset, with an eccentricity of 0.11, shows that \texttt{AWaRe} can closely follow the injected waveform, with the reconstruction nearly indistinguishable from the original signal. In contrast, the bottom inset, with an eccentricity of 0.31, while exhibiting more pronounced deviations, still demonstrates the model's ability to capture the essential features of the waveform. The gray lines represent the whitened strain data, the red lines indicate the whitened injected waveforms, and the blue bands depict the 90\% C.I. of the reconstructions. These insets highlight the model's capacity to handle the modulation in amplitude and phase introduced by non-zero eccentricities. \\

The results presented here underscore \texttt{AWaRe}'s robust generalization capabilities beyond its training set. The model's ability to accurately reconstruct waveforms with low to moderate eccentricities, and still produce reliable results at higher eccentricities, demonstrates its potential for broader applicability in gravitational wave data analysis. This robustness is particularly important for detecting and analyzing signals from sources with significant eccentricities, which provide valuable insights into the formation and evolution of compact binary systems.



\section{Conclusion}

In this paper, we demonstrated the robustness and generalizability of \texttt{AWaRe} for reconstructing GW signals from BBH mergers, including signals with higher-order modes and eccentricities. Building on our previous study, we have incorporated waveform reconstruction uncertainty in the our deep learning framework, enhancing the robustness of its predictions. Our tests, conducted on a diverse set of injections with parameters beyond the training set, illustrate \texttt{AWaRe}'s capacity to handle a variety of complex waveform features. Specifically, the model was able to reconstruct waveforms with higher masses, significant mass ratios, varied waveform systematics and non-zero eccentricities, despite being trained exclusively on non-eccentric waveforms with much smaller primary masses and a fixed waveform approximant. \\

The ability of \texttt{AWaRe} to generalize beyond its training set has significant implications for GW data analysis. This robustness enhances its utility in real-world scenarios, where the variety of source parameters can be vast and unpredictable. By accurately reconstructing waveforms with complex features not present in the training set, \texttt{AWaRe} opens new avenues for the detection and analysis of GW signals from diverse astrophysical sources. This generalization capability is crucial for advancing our understanding of the universe's most extreme events and improving current GW search methodologies. \\

We plan to extend this work by adapting \texttt{AWaRe} into a full GW search and reconstruction pipeline, operating at low latency. This is motivated by \texttt{AWaRe}'s demonstrated ability to distinguish between actual GW signals and pure noise, as shown in our previous work \citep{AWaRe}. The robustness of our model to outliers from the source parameters' prior distributions further builds confidence in its applicability for GW detection in future observation runs. Additionally, \texttt{AWaRe} is uniquely suited to produce rapid reconstructions of a wide variety of GW transients. While cWB and BayesWave reconstructions are mostly optimized for burst-like sources, \texttt{AWaRe} can generate reconstructions at millisecond latency for signals originating from a much wider range of BBH masses. We also plan to perform robust tests of \texttt{AWaRe} against glitches in our future work. This study is especially important to characterize \texttt{AWaRe}'s ability to accurately distinguish real GW signals from noise transients and perform high-fidelity reconstructions of astrophysical waveforms. Furthermore, training \texttt{AWaRe} on synthetic glitches sampled from a wide range of glitch classes will make it suitable for application in any detector characterization framework and complement online GW search methods.

\begin{acknowledgments}
The authors would like to thank Thomas Dent and Alistair McLeod for their helpful comments and suggestions. This research was undertaken with the support of compute grant and resources, particularly the DGX A100 AI Computing Server, offered by the Vanderbilt Data Science Institute (DSI) located at Vanderbilt University, USA. This research used open public data obtained from the Gravitational Wave Open Science Center (https://www.gw-openscience.org), a service of LIGO Laboratory, the LIGO Scientific Collaboration and the Virgo Collaboration. LIGO is funded by the U.S. National Science Foundation. Virgo is funded by the French Centre National de Recherche Scientifique (CNRS), the Italian Istituto Nazionale della Fisica Nucleare (INFN) and the Dutch Nikhef, with contributions by Polish and Hungarian institutes. This material is based upon work supported by NSF's LIGO Laboratory which is a major facility fully funded by the National Science Foundation. 
\end{acknowledgments}

\bibliography{sample631}{}
\bibliographystyle{aasjournal}



\end{document}